\documentclass[11pt,a4paper]{article}
\usepackage[hyperref]{emnlp2018}
\usepackage{times}
\usepackage{latexsym}
\usepackage{amsmath}

\usepackage{url}

\usepackage[utf8]{inputenc}

\usepackage{color}
\usepackage{rotating}
\usepackage{booktabs} % For formal tables
\usepackage{graphicx}
\usepackage{subcaption}
\usepackage{mwe}
\usepackage{multirow}
\usepackage{bbm}
\usepackage{enumerate}
\usepackage{enumitem}

\aclfinalcopy % Uncomment this line for the final submission
\title{The Remarkable Benefit of User-Level Aggregation for Lexical-based Population-Level Predictions}
\author{Salvatore Giorgi$^1$  Daniel Preo$ţ$iuc-Pietro$^2$ {\bf Anneke Buffone}$^1$ \\
{\bf Daniel Rieman}$^1$  {\bf Lyle H. Ungar}$^2$  \and {\bf H. Andrew Schwartz}$^3$ \\
$^1$Department of Psychology, University of Pennsylvania\\
$^2$Computer and Information Science, University of Pennsylvania\\
$^3$Computer Science, Stony Brook University\\
{\tt sgiorgi@sas.upenn.edu}}

\date{}

\begin{document}
\maketitle
\begin{abstract}

Nowcasting based on social media text promises to provide unobtrusive and near real-time predictions of community-level outcomes. These outcomes are typically regarding \textit{people}, but the data is often aggregated without regard to users in the Twitter populations of each community. This paper describes a simple yet effective method for building community-level models using Twitter language aggregated by user. Results on four different U.S. county-level tasks, spanning demographic, health, and psychological outcomes show large and consistent improvements in prediction accuracies (e.g. from Pearson $r=.73$ to $.82$ for median income prediction or $r=.37$ to $.47$ for life satisfaction prediction) over the standard approach of aggregating all tweets. We make our aggregated and anonymized community-level data, derived from 37 billion tweets -- over 1 billion of which were mapped to counties, available for research.
\end{abstract}

\section{Introduction}

Social media is an increasingly popular resource for large-scale population assessment which promises a cheap and non-intrusive complement to standard surveys with finer spatio-temporal scales~\cite{coppersmith2015adhd,mowery2016towards,wang2017detecting}. 
Twitter has been used --- among other things --- to measure community health~\cite{paul2011you,mowery2016towards, eichstaedt2015psychological}, well-being~\cite{schwartz2013characterizing}, and public opinion on politics~\cite{o2010tweets,miranda2015twitter}. By having access to measurements from multiple locations or communities, models trained on text data from social media can be used both to predict future measurements and to provide community estimates where these are lacking or are not robust. Such research is made possible by the massive amount of easily accessible user-generated data from public social media. 

% we could also mention some early forecasting papers (like Brendan O'Connor with politics, or Bollen with stock prices, Bill Lampos with flu rates where they aggregated all tweets in a time range, although their experimental setup was over time rather than over space

However, there has been little research on the way in which such data should be aggregated in order to compute community-level lexical feature estimates. 
Typically, data are aggregated in a ``bag of words'' style, disregarding tweets and authors~\cite{culotta2014estimating,schwartz2013characterizing,eichstaedt2015psychological,curtis}. We find, however, that giving equal weight to each user, rather than to each word or tweet, yields much more accurate community-level predictions.

%A: does it matter that it's the younger group that posts more? This may bring up selection bias. 
%For example, accounts of younger users may post more content as these users tend to use Twitter more as a medium for communication and may skew the community estimates. 
%Building methods that take into account the user-level of analysis have the promise to alleviate these downsides and improve predictive performance and interpretability.

In this paper, we conduct a series of experiments testing various simple yet intuitive aggregation methods. We show that choice of aggregation methods can result in substantial (one might even say ``remarkable'') boosts in accuracy when predicting U.S. county level outcomes (e.g. user-to-county aggregation yields a $7\%$ to $27\%$ increase in Pearson correlation).
Contributions include (a) validation of aggregation approaches across four outcomes related to health, psychology, and demographics, (b) validation that aggregation has some effect on smaller sample of Twitter data, (c) show the effect of power tweeters (or ``super users") and (d) release of resource-intensive community aggregated lexical data.

\paragraph*{Related work.}
This is the first work we know of to explore simple aggregation techniques for population-level prediction tasks from language. 
Previous work has explored more sophisticated adjustments, such as addressing demographic-self selection bias in Twitter community predictions by re-weighting messages, finding small improvements (a 4.5\% reduction in symmetric mean absolute percentage error)~\cite{culotta14}. 
%. This showed a slight improvement in However, none have considered the use user-level information in aggregation.
%Other have 
In a political voting intention prediction application,   ~\cite{lampos13}
modeled users and words jointly by learning separate regression weights for the two dimensions based on the intuition that each user contributes differently towards the outcome. However, their model was specifically adapted to problems that use time-series outcomes, rather than community-level aggregation.
Distributions of lexical features are considered at multiple levels of analysis (message, user and community) in \cite{lexfeatdist17acl} though each level considers one type of aggregation. 
Similar aggregation methods have been used in the context of topic modeling (Latent Dirichlet Allocation~\cite{blei2003latent} and Author Topic
Model~\cite{rosen2004author}) by considering user, hashtag and conversation level aggregations~\cite{alvarez2016topic,hong2010empirical} but, again, community level aggregation and prediction tasks were not considered.

\section{Data}
Research was reviewed by an academic institutional review board and deemed exempt.

\subsection{Twitter Data Collection}

\paragraph*{Twitter Sample}
A random 10\% sample of the entire Twitter stream (`GardenHose') was collected between July 2009 and April 2014, which was then supplemented with a random 1\% sample from May 2014 to February 2015. The total sample contains approximately 37.6 billion tweets \cite{preotiuc2012trendminer}.

\paragraph*{County Mapping}
In order to map each tweet to a location within a county in the United States, we use both self-reported location information in user profiles and latitude/longitude coordinates associated with a tweet. If latitude/longitude coordinates are present then we trivially map the tweet to a county. The self-reported location information is a free text field and we use a cascading set of rules to map this field to a county. The rules are designed to avoid false positives (incorrect mappings) at the expense of fewer mappings. The full details of this process can be found in~\cite{schwartz2013characterizing}. Note that the latitude/longitude coordinates are a tweet attribute whereas the self-reported location is a user attribute yet both are used to map tweets to counties. Users are assigned a county by considering their earliest county mapped tweet. 

In total, we are able to map 1.78 billion of the 37.6 billion tweets to a US county using the above-mentioned method. The county mapped data set was then filtered to contain only English tweets using the popular langid.py method~\cite{lui2012langid}, further reducing our tweet set to 1.64 billion tweets. For experiments with user-level data aggregation, we removed users who made relatively few (less than 30) posts in our data set.

\paragraph{Publicly Available Stream}
The standard publicly available Twitter stream outputs approximately 1\% of the public Tweet volume. Since a 10\% sample is not available to most researchers, we replicated a 1\% sample by taking a random 10\% of our county mapped, English filtered 10\% sample. The same process of county mapping, language filtering and user selection was applied to this data resulting in 131 million county mapped English tweets from 1.57 millions users. Table~\ref{table:data} presents the data set statistics.

\paragraph{The County Tweet Lexical Bank}
The County Tweet Lexical Bank is a U.S. County level data set comprised of two feature sets\footnote{Available at \url{https://github.com/wwbp}} : 
\begin{itemize}[noitemsep,topsep=0pt,leftmargin=*]
\item an aggregated ``bag-of-words" count vector across all the county's messages in order to preserve anonymity. The unigrams represent the most frequent words in the data set;\footnote{While 25,000 features were used in the predictive tasks we removed some features (@-mentions, URLs, etc.) from the data release to preserve anonymity.} ;
\item a ``bag-of-topics" representation for each county, with 2000 social media-derived topics described in~\cite{schwartz2013characterizing}.
\end{itemize}
Both feature sets will be releases across the 2009-2015 time span as well as individual years. Yearly updates will be included as they become available. As we are only releasing aggregated word-level features, as opposed to raw Tweets, this data release is within Twitter's Terms of Service.

\begin{table}[]
\centering
\resizebox{\columnwidth}{!}{
    \begin{tabular}{cccccc}
    \toprule
    \multirow{2}{*}{} & \multicolumn{4}{c}{Number of Tweets} & \multirow{2}{*}{\begin{tabular}[c]{@{}c@{}}Number\\ of\\ Users\end{tabular}} \\ \cline{2-5}
     & \begin{tabular}[c]{@{}c@{}}Full\\ Sample\end{tabular} & \begin{tabular}[c]{@{}c@{}}County\\ Mapped\end{tabular} & English$^*$ & \begin{tabular}[c]{@{}c@{}}User\\ Level$^{**}$\end{tabular} &  \\ \hline
    10\% & 37.6B & 1.78B & 1.64B & 1.53B & 5.25M \\ 
    1\% & \textbf{-} & - & 199M & 131M & 1.57M \\ \toprule
    \end{tabular}
}
\caption{Number of tweets in each section of the resource, including the total number of users. (*) The number of tweets used in the ``all" experiments; (**) the number of tweets in the remaining experiments.}
\label{table:data}
\end{table}

\subsection{Outcomes}

The following U.S. county demographic, psychological and health variables were used in our prediction tasks. Table \ref{table:county data} gives statistics for each county variable. 

\paragraph{Income and Education}
The census data for county median household income (log-transformed to reduce skewness; $N$=1,750) and percentage of people with a Bachelor's degree ($N$=1,750) were obtained from the 2010 U.S. Census Bureau's American Community Survey (ACS).

\paragraph{Life Satisfaction} To assess subjective well-being we used the average response to the question “In general, how satisfied are you with in your life?” (1 = very dissatisfied and 5 = very satisfied)~\cite{lawless2011predictors}. Estimates are averaged across 2009 and 2010 ($N$=1,952).

\paragraph{Mortality Rates}
From the Centers of Disease Control and Prevention (CDC) we collected age-adjusted mortality rates for heart disease ($N$=2,041). Rates are averaged across 2010-2015.

\begin{table}[]
\resizebox{\columnwidth}{!}{\begin{tabular}{lcccccc}\toprule
                                                         & N & Mean & Std Dev & Min & Max & Skew \\ \hline
Income & 1750 & 4.66 & 0.11 & 4.33 & 5.07 & 0.47 \\ 
Educat. & 1750 & 21.57 & 9.46 & 5.70 & 70.30 & 1.20 \\ 
Life Satis. & 1952 & 3.39 & 0.03 & 3.26 & 3.51 & 0.02 \\ 
Heart Dis. & 2041 & 186.66 & 45.59 & 54.82 & 412.32 & 0.66   \\  
\bottomrule\end{tabular}}
\caption{Descriptives of U.S. County data used in the four prediction tasks.}
\label{table:county data}
\end{table}

\section{Methods}

\subsection{Aggregation}

Our aim is to use the user-level information based on the assumption that aggregating data first at the user-level would remove biases introduced by non-standard users of the platform. 
To this end, we explore three types of aggregation: (1) tweet to county, (2) county ``bag of words" and (3) user to county.

\paragraph{Tweet to County} 
Here we compute
\begin{equation}
    \text{feat}_{i,j} = \frac{1}{N_j}\sum_{k}\mathbbm{1}_i (\text{unigram}_{k}),
\label{eq:tweet level extraction}
\end{equation}
where $\mathbbm{1}_i$ denotes the indicator function for unigram$_i$. Here the $i$th feature for the $j$th unit of analysis (a U.S. county) is equal to the relative frequency of the unigram: the number of times each unigram was mentioned divided by $N_j$ the total number of tweets from county $j$. 

\paragraph{County} Next, we use a method which was generally used in past research, which aggregates all messages to a community disregarding any meta-data, including tweet or user information.
Previous state-of-the-art results using this method include life satisfaction $r$ = .31 \cite{schwartz2013characterizing}, atheroclerotic heart disease $r$ = .42~\cite{eichstaedt2015psychological} and education $r$ = 0.15~\cite{culotta14}.
We therefore consider each county a ``bag of words" using (\ref{eq:tweet level extraction}) with $N_j$ equal to the number of unigrams from county $j$. 

\paragraph{User to County} The third method treats the unit of analysis (U.S. county) as a community of users. Therefore, feature weights are extracted at the user level, normalized and then averaged to communities:

\begin{equation}
    \text{feat}_{i,j} = \frac{1}{N_j}\sum_{k\in U_j}r_k(\text{unigram}_{i}),
\label{eq:feat agg}
\end{equation}
where $U_j$ is the set of users in county $j$, $N_j$ is the total number of Twitter users in county $j$ and $r_k(x)$ is the relative frequency of feature $x$ for user $k$ with $i\in \lbrace$all unigrams$\rbrace$ and $j\in \lbrace$all counties$\rbrace$.

\subsection*{Features}
We use as features a list of 2,000 social media-derived topics generated from Latent Dirichlet Allocation~\cite{blei2003latent} using the complete MyPersonality Facebook data set consisting of approximately 15 million posts~\cite{schwartz2013characterizing}. The topic loadings are computed from the most frequent 25,000 unigrams in our data set. We also use a subset of these unigrams as additional features in our models (25,000 reduced to 10,000). 

\subsection*{Experimental setup}

For each of the four county level Census and health variables we built three models using 10-fold cross validation with the following features: (1) unigrams, (2) topics and (3) unigrams + topics. For consistency across tasks we only considered counties with 100 or more 30+ tweet users ($N$=2,041). 

We used a feature selection pipeline which first removed all low variance features and then features that were not correlated with our census and health data. % (using a family-wise error rate alpha of 60 \lhu{you mean 60\% ?}).
Principal component analysis was then applied to the reduced feature set for further dimensionality reduction. This preprocessing was used to avoid overfitting, since our model included more independent variables (2000 topic frequencies and/or 10k unigrams) than observations (at most 2,041 counties). For the prediction task we used linear regression with $\ell_2$ regularization (Ridge regression)~\cite{eichstaedt2015psychological}. The regression regularization parameter $\alpha$ was set to 1000 using grid search.

Because our initial dataset consisted of 37.6 billion tweets, using distributed IO was crucial for data aggregation and feature extraction. 
We used a Hadoop-style cluster consisting of 64 disks and 64 CPU cores across 8 physical machines. 
Over this cluster, we used Hadoop MapReduce for the county mapping step (taking approximately 1 week of wall clock runtime) and Spark for the feature aggregations (taking approximately 1 day of wall clock runtime). 
The entire pipeline of county mapping, English language filtering, feature extraction and prediction used the DLATK Python package~\cite{schwartz2017dlatk}\footnote{Available at \url{https://github.com/dlatk}}.

\begin{table}[t]
    \centering
  \begin{subtable}{\columnwidth}
        \centering
        \resizebox{\columnwidth}{!}{
            \begin{tabular}{lcccccccc}\toprule
            \multirow{2}{*}{}  & \multicolumn{1}{c}{Income} & \multicolumn{1}{c}{Educat.} & \multicolumn{1}{c}{\begin{tabular}[c]{@{}c@{}}Life \\ Satis.\end{tabular}} & \multicolumn{1}{c}{\begin{tabular}[c]{@{}c@{}}Heart \\ Disease\end{tabular}} \\ \hline
            Tweet to County   & .68  & .80 & .26 & .70 \\  
            County            &  .73 & .80 & .37 & .70\\  
            %County (all)       &  .73 & .83 & .31 & .72\\ 
            User to County &  \textbf{.82} & \textbf{.88} & \textbf{.47} & \textbf{.75}\\ \bottomrule 
            \end{tabular}
        }
        \caption{ Unigrams + Topics, Pearson $r$}
    \end{subtable}%
    \hfill     
    \begin{subtable}{\columnwidth}
        \centering
        \resizebox{\columnwidth}{!}{
            \begin{tabular}{lcccccccc}\toprule
            \multirow{2}{*}{}  & \multicolumn{1}{c}{Income} & \multicolumn{1}{c}{Educat.} & \multicolumn{1}{c}{\begin{tabular}[c]{@{}c@{}}Life \\ Satis.\end{tabular}} & \multicolumn{1}{c}{\begin{tabular}[c]{@{}c@{}}Heart \\ Disease\end{tabular}} \\ \hline
            Tweet to County   & .67  & .79 & .22 &  .65 \\  
            County            & .72 & .78 & .37 & .64 \\  
            %County (all)       & .74 & .83 & .30 & .69 \\ 
            User to County & \textbf{.79} & \textbf{.87} & \textbf{.44} & \textbf{.73}  \\ \bottomrule
            \end{tabular}
        }
        \caption{Unigrams, Pearson $r$}
    \end{subtable}%
    \hfill  
    \begin{subtable}{\columnwidth}
        \centering
        \resizebox{\columnwidth}{!}{
            \begin{tabular}{lcccccccc}\toprule
            \multirow{2}{*}{}  & \multicolumn{1}{c}{Income} & \multicolumn{1}{c}{Educat.} & \multicolumn{1}{c}{\begin{tabular}[c]{@{}c@{}}Life \\ Satis.\end{tabular}} & \multicolumn{1}{c}{\begin{tabular}[c]{@{}c@{}}Heart \\ Disease\end{tabular}} \\ \hline
            Tweet to County   & .65 & .77 & .31 & .71  \\  
            County           & .68 & .80 & .34 & .71  \\ 
            %County (all)       & .68 & .81 & .33 & .72  \\ 
            User to County &  \textbf{.81} & \textbf{.87} & \textbf{.47} & \textbf{.76} \\ \bottomrule
            \end{tabular}
        }
        \caption{Topics, Pearson $r$}
    \end{subtable}%

    \caption{Prediction results (Pearson $r$) for direct aggregation comparison on the 10\% sample.}
    \label{table: 10pct prediction}
\end{table}

\begin{table}[t]
        \centering
        \resizebox{\columnwidth}{!}{
            \begin{tabular}{lcccccccc}\toprule
            \multirow{2}{*}{}  & \multicolumn{1}{c}{Income} & \multicolumn{1}{c}{Educat.} & \multicolumn{1}{c}{\begin{tabular}[c]{@{}c@{}}Life \\ Satis.\end{tabular}} & \multicolumn{1}{c}{\begin{tabular}[c]{@{}c@{}}Heart \\ Disease\end{tabular}} \\ \hline
            User to County &  \textbf{.82} & \textbf{.88} & \textbf{.47} & \textbf{.75}\\ 
            $N_{user-tweets}$ & 1.350B & 1.350B & 1.356B & 1.360B  \\ \hline
            
            Tweet to County (all)   & .72 & .81 & .36 & .71 \\  
            County (all)       &  .73 & .82 & .31 & .72\\ 
            $N_{all-tweets}$ & 1.621B & 1.621B & 1.628B & 1.634B  \\ 
            
            \bottomrule
            \end{tabular}
        }
        \caption{Prediction results (Pearson $r$, using unigrams + topics) using full 10\% data vs. users with 30+ tweets. The number of tweets used in each task is listed to highlight the fact that the ``User to County" tasks use less tweets than the ``all" tasks. }
    \label{table: 10pct all prediction}
    \end{table}

\subsection*{Experiments}

Using the above setup we perform 3 experiments in order to explore the effects of data aggregation. We 1) directly compare aggregation methods using our 10\% data; 2) compare aggregation methods using a 1\% sample and, finally, 3) explore the effect of choosing an upper bound on the number of posts per Twitter users, looking at users with less than 50, 500, 1000 posts. This allows us to exclude frequent posters who are potentially organizations or bots.

\section{Results and Discussion}

\paragraph{Direct aggregation comparison.}  The results of our predictive experiments on the 10\% data can be found in Table \ref{table: 10pct prediction}. Across all four tasks we see that the ``User to County" approach outperforms the other aggregation methods, giving a higher Pearson $r$ and obtaining state-of-the-art results for community-level predictions.

We see the largest gains for the ``User to County" aggregation for the income outcome, with a 13 point increase in Pearson $r$ for topics alone and a 9 point increase for unigrams + topics.

In Table \ref{table: 10pct all prediction} we remove the 30+ tweet requirement from the ``Tweet to County" and ``County" methods and compare against the ``User to County" method (with the 30+ tweet requirement). Again we see the ``User to County" method outperforms all others in spite of the fact that the ``User to County" approach uses less data than both ``all" approaches, which contains ~108 million more tweets.

\begin{table*}[t]
\centering
\resizebox{4.8in}{!}{\begin{tabular}{lcccccccc}\toprule
 & \multicolumn{2}{c}{Income} & \multicolumn{2}{c}{Educat.} & \multicolumn{2}{c}{\begin{tabular}[c]{@{}c@{}}Life \\ Satis.\end{tabular}} & \multicolumn{2}{c}{\begin{tabular}[c]{@{}c@{}}Heart \\ Disease\end{tabular}} \\ \hline
Tweet to County  & .71 & .62 & .77 & .71 & .35 & .32 & .64 & .63 \\
County & .70 & .60 & .76 & .67 & .32 & .28 & .62 & .62 \\
User to County & .76 & .70 & .79 & .74 & .39 & .28 & .66 & .66 \\
$N_{user-tweets}$ & 127M & 130M & 127M & 130M & 127M & 130M & 127M & 131M \\ \hline
County (all) & .75 & .67 & .83 & .77 & .37 & .34 & .68 & .66 \\ 
$N_{all-tweets}$ & 191M & 195M & 191M & 195M & 191M & 197M & 191M & 198M \\ \hline
$N_{counties}$ & 949 & 1750$^*$ & 949 & 1750$^*$ & 954 & 1952$^*$ & 960 & 2041$^*$ \\
\bottomrule
\end{tabular}}
\caption{1\% sample prediction results (Pearson $r$) using topics + unigrams. $*$ same counties as the 10\% prediction task.}
\label{table:1pct vs 10pct}
\end{table*}

\begin{table}[]
\centering
\resizebox{\columnwidth}{!}{\begin{tabular}{ccccccc}\toprule
& \begin{tabular}[c]{@{}c@{}}Max \\ Tweets\end{tabular} & Income    & Educat.   & \begin{tabular}[c]{@{}c@{}}Life \\ Satis.\end{tabular} & \begin{tabular}[c]{@{}c@{}}Heart \\ Disease\end{tabular} & \begin{tabular}[c]{@{}c@{}}Num. Users \\ Removed\end{tabular} \\ \hline
\multirow{4}{*}{\rotatebox[origin=c]{90}{\begin{tabular}[c]{@{}c@{}}County\\ (all)\end{tabular}}}& 50     & .73 & .84 & .34 & .68 & 4,665,114 \\
%& 100    &  &  &  &  & 2,961,252 \\
& 500    & .81 & .87 & .44 & .75 & 611,661 \\
& 1000   & .81 & .87 & .41 & .75 & 217,517 \\ 
& No Max & .73 & .82 & .31 & .72 & - \\ \hline
\multirow{4}{*}{\rotatebox[origin=c]{90}{\begin{tabular}[c]{@{}c@{}}User to\\ County\end{tabular}}}& 50     & .68 & .80 & .34 & .64 & 4,665,114 \\
%& 100    & .76 & .85 & .42 & .72 & 2,961,252 \\
& 500    & .80 & .87 & .47 & .76 & 611,661 \\
& 1000   & .81 & .87 & .47 & .76 & 217,517 \\ 
& No Max & .81 & .87 & .48 & .76 & - \\ \bottomrule                                                           
\end{tabular}}
\caption{Prediction results (Pearson $r$) using topics + unigrams. Users with more than ``Max Tweets" number of tweets are removed from the sample.}
\label{table:super-super users}
\end{table}

\paragraph{1\% data.} In Table \ref{table:1pct vs 10pct} we repeat the above experiment on a 1\% Twitter sample. Here we see that the ``User to County" method outperforms both the ``Tweet to County" and ``County" methods (with all three tasks using the same number of tweets). When we compare the ``User to County" and ``County (all)" methods we see the ``User to county" outperforming on two out of four tasks (Income and Life Satisfaction). Again, we note that the ``User to County" is using less data than the ``County (all)". While, across the board, the performance increase is not as substantial as in the 10\% results, we see comparable performance between ``User to County" and ``County (all)" methods despite the difference in the number of tweets.

\paragraph{Super users.} 
One theory why we see such large gains depending on aggregation technique is that aggregating through users negates the effects of super users -- those who post an extraordinary amount (such as organizations or bots). We implemented a maximum tweet requirement in order to remove these users and see if that accounts for the difference. Here we use both the ``User to County" and ``County (all)" samples and report results in Table \ref{table:super-super users}. These results demonstrate that by keeping only users with less than 500 tweets we get results close to our ``User to County (No Max)" method using the user-naive ``County (all)" scheme. This shows that relatively few users (in this case 611k) can significantly decrease performance, though still leaves a small gain from the user to county approach. 
As seen in the lower half of the table, this thresholding does not increase performance when using the ``User to County" method, which suggests such users can still be beneficial if they are just treated such that they can't dominate a community. 
This highlights the benefit of our simple method: we do not need to consider optimizations which may not generalize across data, such as upper-bound thresholds on the number of tweets per user. Further, the user-to-county aggregation seems to provide at least a small benefit beyond removal of super users.

\section{Conclusion}

This study explored the benefit of aggregation techniques for streaming user-generated data from individual messages to community level data, the typical setting for nowcasting. We showed that by simply aggregating to users first and then taking the mean within a county, we can obtain large gains (remarkably, up to a 13 point increase in Pearson correlation) over typical aggregation methods common in past work. In order to foster nowcasting research utilizing this more ideal aggregation, we will release the County Tweet Lexical Bank -- a large aggregated and anonymized county-level data set, and computed on more than 1.6 billion tweets posted over 5 years.

Future work in this area can look at adjusting models to account for other meta-data such as temporal variation and diversity and to adjust for selection biases present in social media, where the user base on social media is not representative of the population of the community~\cite{pew16}.

\section*{Acknowledgments}
This work was supported by a grant from the Templeton Religion Trust (ID \#TRT0048). The funders had no role in study design, data collection and analysis, decision to publish, or preparation of the manuscript. 
% The acknowledgments should go immediately before the references.  Do
% not number the acknowledgments section. Do not include this section
% when submitting your paper for review. \\

\bibliographystyle{acl_natbib_nourl}
\bibliography{emnlp2018}

\end{document}

\maketitle

\section{Appendix A. Additional Experiments and Results}

% \paragraph{County and Tweet Counts} Table \ref{table: sample size} shows the number of tweets and counties used in each task: $N_{tweets}$ is the number of tweets used in the ``Tweet to County", ``County" and ``User to County" aggregations and $N_{ALLtweets}$ is the number of tweets used in the ``County (all)" tasks. 

% \begin{table}{\columnwidth}
%     \resizebox{1.\columnwidth}{!}{
%         \begin{tabular}{lcccccccc}\toprule
%         \multirow{2}{*}{}  & \multicolumn{1}{c}{Income} & \multicolumn{1}{c}{Educat.} & \multicolumn{1}{c}{\begin{tabular}[c]{@{}c@{}}Life \\ Satis.\end{tabular}} & \multicolumn{1}{c}{\begin{tabular}[c]{@{}c@{}}Heart \\ Disease\end{tabular}}  \\ \hline
%         $N_{counties}$ & 1750 & 1750 & 1952 & 2041  \\ 
%         $N_{tweets}$  & 1.350B & 1.350B & 1.356B & 1.360B \\ \bottomrule
%         $N_{ALLtweets}$ & 1.621B & 1.621B & 1.628B & 1.634B  \\ \bottomrule
%         \end{tabular}
%     }
%     \caption{Number of counties and tweets}
%     \label{table: sample size}
% \end{table}

\paragraph{Direct aggregation comparison.}  In addition to the Pearon $r$ results above we report Mean Squared Error (MSE) in Table \ref{table: 10pct prediction mse} 

\begin{table}[htb]
    \centering
%   \begin{subtable}{\columnwidth}
%         \centering
%         \resizebox{\columnwidth}{!}{
%             \begin{tabular}{lcccccccc}\toprule
%             \multirow{2}{*}{}  & \multicolumn{1}{c}{Income} & \multicolumn{1}{c}{Educat.} & \multicolumn{1}{c}{\begin{tabular}[c]{@{}c@{}}Life \\ Satis.\end{tabular}} & \multicolumn{1}{c}{\begin{tabular}[c]{@{}c@{}}Heart \\ Disease\end{tabular}} \\ \hline
%             Tweet to County   &   &  &  &  \\  
%             County            &  .73 & .80 & .37 & .70\\  
%             %County (all)       &  .73 & .83 & .31 & .72\\ 
%             User to County &  \textbf{.82} & \textbf{.88} & \textbf{.47} & \textbf{.75}\\ \bottomrule 
%             \end{tabular}
%         }
%         \caption{ Unigrams + Topics, Mean Squared Error (MSE)}
%     \end{subtable}%
%     \hfill
    \begin{subtable}{\columnwidth}
        \resizebox{\columnwidth}{!}{
            \begin{tabular}{lcccccccc}\toprule
            \multirow{2}{*}{}  & \multicolumn{1}{c}{Income} & \multicolumn{1}{c}{Educat.} & \multicolumn{1}{c}{\begin{tabular}[c]{@{}c@{}}Life \\ Satis.\end{tabular}} & \multicolumn{1}{c}{\begin{tabular}[c]{@{}c@{}}Heart \\ Disease\end{tabular}} \\ \hline
            Tweet to County   & 6.67e-3 & 32.9 & 8.81e-4 & 1083 \\ 
            County           & 5.49e-3 & 35.1 & 6.49e-4 & 1105 \\ 
            User to County & \textbf{3.68e-3}  & \textbf{20.7} & \textbf{5.51e-4} & \textbf{913}\\ \bottomrule
            \end{tabular}
        }
        \caption{Unigrams + Topics, Mean Squared Error (MSE)}
    \end{subtable}% 
    \hfill     
    \begin{subtable}{\columnwidth}
        \centering
        \resizebox{\columnwidth}{!}{
            \begin{tabular}{lcccccccc}\toprule
            \multirow{2}{*}{}  & \multicolumn{1}{c}{Income} & \multicolumn{1}{c}{Educat.} & \multicolumn{1}{c}{\begin{tabular}[c]{@{}c@{}}Life \\ Satis.\end{tabular}} & \multicolumn{1}{c}{\begin{tabular}[c]{@{}c@{}}Heart \\ Disease\end{tabular}} \\ \hline
            Tweet to County   & 6.76e-3 & 35.1 & 1.02e-3 & 1289  \\  
            County            & 6.44e-3 & 34.3 & 6.53e-4 & 1232 \\  
            %County (all)       & .74 & .83 & .30 & .69 \\ 
            User to County & \textbf{4.18e-3} & \textbf{22.9} & \textbf{5.75e-4} & \textbf{1028}  \\ \bottomrule
            \end{tabular}
        }
        \caption{Unigrams, Mean Squared Error (MSE)}
    \end{subtable}%
    \hfill  
    \begin{subtable}{\columnwidth}
        \centering
        \resizebox{\columnwidth}{!}{
            \begin{tabular}{lcccccccc}\toprule
            \multirow{2}{*}{}  & \multicolumn{1}{c}{Income} & \multicolumn{1}{c}{Educat.} & \multicolumn{1}{c}{\begin{tabular}[c]{@{}c@{}}Life \\ Satis.\end{tabular}} & \multicolumn{1}{c}{\begin{tabular}[c]{@{}c@{}}Heart \\ Disease\end{tabular}} \\ \hline
            Tweet to County   & 6.73e-3 & 38.0 & 6.61e-4 & 1034  \\  
            County           & 6.39e-3 & 32.8 & 6.40e-4 & 1059 \\ 
            %County (all)       & .68 & .81 & .33 & .72  \\ 
            User to County &  \textbf{3.83e-3} & \textbf{21.5} & \textbf{5.28e-4} & \textbf{866} \\ \bottomrule
            \end{tabular}
        }
        \caption{Topics, Pearson $r$}
    \end{subtable}%
    
    \caption{Prediction results, reported Mean Squared Error (MSE). }
    \label{table: 10pct prediction mse}
\end{table}

\paragraph{Users per county.} To gain further insight into differences in accuracy, we look at accuracy as a function of our users per county requirement. For this task we consider the ``User to County" approach and build a models using unigrams + topics, varying the required number of users. Figure~\ref{fig:30 tweet users} shows the results. 
We see a general increase in accuracy as the user-threshold is raised. 
Of course, this happens at the expense of covering fewer counties, varying from 2153 at the 10 user requirement to 626 counties at the 1000 user requirement.

\begin{figure}[h]
    \centering
    \includegraphics[width=\linewidth]{naacl_30tweetusers.pdf}
    \caption{Prediction results (Pearson $r$) using 10\% ``User to County" (unigrams + topics) when varying the minimum number of Twitter users per county. Parenthetical number indicates the average number of counties across tasks after meeting the minimum user threshold.} 
    \label{fig:30 tweet users}
\end{figure}

\paragraph{Replication across time.} Here we explore the effect aggregation has on replication over time, considering separate years of age adjusted heart disease mortality rates and aggregate Twitter data over single years (from 2012 and 2013). We chose heart disease as the other variables (income, education and life satisfaction) are not produced every year. Additionally, we chose 2012 and 2013 as the other years in our Twitter sample contain holes or time periods with a 1\% sample. Results are shown in Table \ref{table: time replication}. Here see the same patterns as previous experiments: ``County" performs better than ``Tweet to County" with ``User to County" outperforming both. We also note that all three methods hold across both years with a slight increase across all tasks in 2013, which contains slightly more data (472 million posts in 2013 vs. 455 million in 2012). 

\begin{table}[]
\centering
\begin{tabular}{lcc}\toprule
                & 2012 & 2013 \\\hline
Tweet to County &  .50  &  .54     \\
County          &  .52  &  .58        \\
User to County  &  \textbf{.58}  &  \textbf{.63}   \\ \bottomrule  
\end{tabular}
\caption{Replication of Heart disease predictions. }
\label{table: time replication}
\end{table}

%\bibliography{emnlp2018}
%\bibliographystyle{acl_natbib_nourl}